\RequirePackage[switch]{lineno_babar_revtex}
\documentclass[twocolumn,aps,prd,superscriptaddress,showpacs]{revtex4}
\catcode`\@=11\relax
\def\@makecol{%
 \setbox\@outputbox\vbox{%
  \boxmaxdepth\@maxdepth
 \protected@write\@auxout{}{%
 \string\@LN@col{\@ifnum{\pagegrid@cur=\@ne}{1}{2}}
      }
  \@tempdima\dp\@cclv
  \unvbox\@cclv
  \vskip-\@tempdima
 }%
 \xdef\@freelist{\@freelist\@midlist}\global\let\@midlist\@empty
 \@combinefloats
 \@combineinserts\@outputbox\footins
  \set@adj@colht\dimen@
  \count@\vbadness
  \vbadness\@M
  \setbox\@outputbox\vbox to\dimen@{%
   \@texttop
   \dimen@\dp\@outputbox
   \unvbox\@outputbox
   \vskip-\dimen@
   \@textbottom
  }%
  \vbadness\count@
 \global\maxdepth\@maxdepth
}%
\def\balance@two#1#2{%
\outputdebug@sw{{\tracingall\scrollmode\showbox#1\showbox#2}}{}%
 \setbox\@ne\vbox{%
  \@ifvoid#1{}{%
   \unvcopy#1\recover@footins
   \@ifvoid#2{}{\marry@baselines}%
  }%
  \@ifvoid#2{}{%
   \unvcopy#2\recover@footins
  }%
 }%
 \dimen@\ht\@ne\divide\dimen@\tw@
 \dimen@i\dimen@
 \vbadness\@M
 \vfuzz\maxdimen
 \loopwhile{%
  \dimen@i=.5\dimen@i
  \outputdebug@sw{\saythe\dimen@\saythe\dimen@i\saythe\dimen@ii}{}%
  \setbox\z@\copy\@ne\setbox\tw@\vsplit\z@ to\dimen@
  \setbox\z@ \vbox{%
 \protected@write\@auxout{}{%
 \string\@LN@col{\@ifnum{\pagegrid@cur=\@ne}{1}{2}}
      }%
   \unvcopy\z@
   \setbox\z@\vbox{\unvbox\z@ \setbox\z@\lastbox\aftergroup\vskip\aftergroup-\expandafter}\the\dp\z@\relax
  }%
  \setbox\tw@\vbox{%
   \unvcopy\tw@
   \setbox\z@\vbox{\unvbox\tw@\setbox\z@\lastbox\aftergroup\vskip\aftergroup-\expandafter}\the\dp\z@\relax
  }%
  \dimen@ii\ht\tw@\advance\dimen@ii-\ht\z@
  \@ifdim{\dimen@i>.5\p@}{%
   \advance\dimen@\@ifdim{\dimen@ii<\z@}{}{-}\dimen@i
   \true@sw
  }{%
   \@ifdim{\dimen@ii<\z@}{%
    \advance\dimen@\tw@\dimen@i
    \true@sw
   }{%
    \false@sw
   }%
  }%
 }%
 \outputdebug@sw{\saythe\dimen@\saythe\dimen@i\saythe\dimen@ii}{}%
\@ifdim{\ht\z@=\z@}{%
\@ifdim{\ht\tw@=\z@}{%
\true@sw
}{%
\false@sw
}%
}{%
\true@sw
}%
{%
}{%
\ltxgrid@info{Unsatifactorily balanced columns: giving up}%
\setbox\tw@\box#1%
\setbox\z@ \box#2%
}%
 \setbox\tw@\vbox{\unvbox\tw@\vskip\z@skip}%
 \setbox\z@ \vbox{\unvbox\z@ \vskip\z@skip}%
 \set@colroom
\dimen@\ht\z@\@ifdim{\dimen@<\ht\tw@}{\dimen@\ht\tw@}{}%
\@ifdim{\dimen@>\@colroom}{\dimen@\@colroom}{}%
 \outputdebug@sw{\saythe{\ht\z@}\saythe{\ht\tw@}\saythe\@colroom\saythe\dimen@}{}%
\setbox#1\vbox to\dimen@{\unvbox\tw@\unskip\raggedcolumn@skip}%
\setbox#2\vbox to\dimen@{\unvbox\z@ \unskip\raggedcolumn@skip}%
\outputdebug@sw{{\tracingall\scrollmode\showbox#1\showbox#2}}{}%
}%
\catcode`\@=12\relax

\usepackage{graphicx}
\usepackage{dcolumn}
\usepackage{amsmath}
\usepackage{epsfig}
\usepackage{diagbox}

\RequirePackage{xspace}

\usepackage{relsize}
\def\babar{\mbox{\slshape B\kern-0.1em{\smaller A}\kern-0.1em
    B\kern-0.1em{\smaller A\kern-0.2em R}}}

\def\epem       {\ensuremath{e^+e^-}\xspace}

\def\Kbar  {\kern 0.2em\overline{\kern -0.2em K}{}\xspace}

\def\Kz    {\ensuremath{K^0}\xspace}
\def\Kzb   {\ensuremath{\Kbar^0}\xspace}
\def\KzKzb {\ensuremath{\Kz \kern -0.16em \Kzb}\xspace}
\def\Kp    {\ensuremath{K^+}\xspace}
\def\Km    {\ensuremath{K^-}\xspace}

\def\KpKm  {\ensuremath{\Kp \kern -0.16em \Km}\xspace}

\def\Dbar    {\kern 0.2em\overline{\kern -0.2em D}{}\xspace}

\def\Dz      {\ensuremath{D^0}\xspace}
\def\Dzb     {\ensuremath{\Dbar^0}\xspace}
\def\DzDzb   {\ensuremath{\Dz {\kern -0.16em \Dzb}}\xspace}
\def\Dp      {\ensuremath{D^+}\xspace}
\def\Dm      {\ensuremath{D^-}\xspace}

\def\DpDm    {\ensuremath{\Dp {\kern -0.16em \Dm}}\xspace}

\def\Bbar    {\kern 0.18em\overline{\kern -0.18em B}{}\xspace}

\def\Bz      {\ensuremath{B^0}\xspace}
\def\Bzb     {\ensuremath{\Bbar^0}\xspace}
\def\BzBzb   {\ensuremath{\Bz {\kern -0.16em \Bzb}}\xspace}
\def\Bu      {\ensuremath{B^+}\xspace}
\def\Bub     {\ensuremath{B^-}\xspace}

\def\BpBm    {\ensuremath{\Bu {\kern -0.16em \Bub}}\xspace}

\def\BorBbar    {\kern 0.18em\optbar{\kern -0.18em B}{}\xspace}
\def\DorDbar    {\kern 0.18em\optbar{\kern -0.18em D}{}\xspace}
\def\KorKbar    {\kern 0.18em\optbar{\kern -0.18em K}{}\xspace}

\mathchardef\Upsilon="7107
\def\Y#1S{\ensuremath{\Upsilon{(#1S)}}\xspace}

\mathchardef\Deltares="7101
\mathchardef\Xi="7104
\mathchardef\Lambda="7103
\mathchardef\Sigma="7106
\mathchardef\Omega="710A

\def\Deltabar{\kern 0.25em\overline{\kern -0.25em \Deltares}{}\xspace}
\def\Lbar{\kern 0.2em\overline{\kern -0.2em\Lambda\kern 0.05em}\kern-0.05em{}\xspace}
\def\Sigbar{\kern 0.2em\overline{\kern -0.2em \Sigma}{}\xspace}
\def\Xibar{\kern 0.2em\overline{\kern -0.2em \Xi}{}\xspace}
\def\Obar{\kern 0.2em\overline{\kern -0.2em \Omega}{}\xspace}
\def\Nbar{\kern 0.2em\overline{\kern -0.2em N}{}\xspace}
\def\Xb{\kern 0.2em\overline{\kern -0.2em X}{}\xspace}

\newcommand{\tev}{\ensuremath{\mathrm{\,Te\kern -0.1em V}}\xspace}
\newcommand{\gev}{\ensuremath{\mathrm{\,Ge\kern -0.1em V}}\xspace}
\newcommand{\mev}{\ensuremath{\mathrm{\,Me\kern -0.1em V}}\xspace}
\newcommand{\kev}{\ensuremath{\mathrm{\,ke\kern -0.1em V}}\xspace}
\newcommand{\ev}{\ensuremath{\mathrm{\,e\kern -0.1em V}}\xspace}
\newcommand{\gevc}{\ensuremath{{\mathrm{\,Ge\kern -0.1em V\!/}c}}\xspace}
\newcommand{\mevc}{\ensuremath{{\mathrm{\,Me\kern -0.1em V\!/}c}}\xspace}
\newcommand{\gevcc}{\ensuremath{{\mathrm{\,Ge\kern -0.1em V\!/}c^2}}\xspace}
\newcommand{\mevcc}{\ensuremath{{\mathrm{\,Me\kern -0.1em V\!/}c^2}}\xspace}

\def\invfb   {\ensuremath{\mbox{\,fb}^{-1}}\xspace}

\def\mus  {\ensuremath{\rm \,\mus}\xspace}

\def\mus        {\ensuremath{\,\mu{\rm s}}\xspace}    

\def\pep2{PEP-II}

\newcommand{\dedx}{\ensuremath{\mathrm{d}\hspace{-0.1em}E/\mathrm{d}x}\xspace}

\def\gsim{{~\raise.15em\hbox{$>$}\kern-.85em
          \lower.35em\hbox{$\sim$}~}\xspace}
\def\lsim{{~\raise.15em\hbox{$<$}\kern-.85em
          \lower.35em\hbox{$\sim$}~}\xspace}

\xspace

\newcommand{\jprlBase}       {Phys.\ Rev.\ Lett.\xspace}
\newcommand{\jprBase}        {Phys.\ Rev.\xspace}
\newcommand{\jplBase}        {Phys.\ Lett.\xspace}
\newcommand{\nimBaseA}       {Nucl.\ Instrum.\ Methods Phys.\ Res., Sect.\ A\xspace}

\newcommand{\cpc}       [1]  {{Comput.\ Phys.\ Commun.\ {\bf #1}}}

\newcommand{\nima}      [1]  {\nimBaseA~{\bf #1}}

\def\evtgen     {\mbox{\tt EvtGen}\xspace}

\def\jetset74   {\mbox{\tt Jetset \hspace{-0.5em}7.\hspace{-0.2em}4}\xspace}

\newcommand{\BABARPubYear}    {14}
\newcommand{\BABARPubNumber}  {013}
\newcommand{\SLACPubNumber} {16212}

\newcommand{\lumi}    {431\invfb}

\def\kk2f       {\mbox{\tt KK2f}\xspace}
\def\tauola     {\mbox{\tt Tauola}\xspace}
\def\photos     {\mbox{\tt Photos}\xspace}
\def\evtgen     {\mbox{\tt EVTGEN}\xspace}
\def\jetset     {\mbox{\tt JETSET}\xspace}

\makeatletter
\newcommand{\thickhline}{%
    \noalign {\ifnum 0=`}\fi \hrule height 1pt
    \futurelet \reserved@a \@xhline
}
\newcolumntype{"}{@{\hskip\tabcolsep\vrule width 1pt\hskip\tabcolsep}}
\makeatother

\newcommand*\patchAmsMathEnvironmentForLineno[1]{%
  \expandafter\let\csname old#1\expandafter\endcsname\csname #1\endcsname
  \expandafter\let\csname oldend#1\expandafter\endcsname\csname end#1\endcsname
  \renewenvironment{#1}%
     {\linenomath\csname old#1\endcsname}%
     {\csname oldend#1\endcsname\endlinenomath}}%
\newcommand*\patchBothAmsMathEnvironmentsForLineno[1]{%
  \patchAmsMathEnvironmentForLineno{#1}%
  \patchAmsMathEnvironmentForLineno{#1*}}%
\AtBeginDocument{%
\patchBothAmsMathEnvironmentsForLineno{equation}%
\patchBothAmsMathEnvironmentsForLineno{align}%
\patchBothAmsMathEnvironmentsForLineno{flalign}%
\patchBothAmsMathEnvironmentsForLineno{alignat}%
\patchBothAmsMathEnvironmentsForLineno{gather}%
\patchBothAmsMathEnvironmentsForLineno{multline}%
}

\begin{document}


\begin{flushleft}
{\babar-PUB-\BABARPubYear/\BABARPubNumber}\\ 
{SLAC-PUB-\SLACPubNumber} 
\end{flushleft}
\title{
{\large \bf \boldmath Measurement of the branching fractions of 
the radiative leptonic $\tau$ decays\\ $\tau \rightarrow e \gamma \nu \bar \nu$ 
and $\tau \rightarrow \mu \gamma \nu \bar \nu$  at \babar}
}

%
\author{J.~P.~Lees}
\author{V.~Poireau}
\author{V.~Tisserand}
\affiliation{Laboratoire d'Annecy-le-Vieux de Physique des Particules (LAPP), Universit\'e de Savoie, CNRS/IN2P3,  F-74941 Annecy-Le-Vieux, France}
\author{E.~Grauges}
\affiliation{Universitat de Barcelona, Facultat de Fisica, Departament ECM, E-08028 Barcelona, Spain }
\author{A.~Palano$^{ab}$ }
\affiliation{INFN Sezione di Bari$^{a}$; Dipartimento di Fisica, Universit\`a di Bari$^{b}$, I-70126 Bari, Italy }
\author{G.~Eigen}
\author{B.~Stugu}
\affiliation{University of Bergen, Institute of Physics, N-5007 Bergen, Norway }
\author{D.~N.~Brown}
\author{L.~T.~Kerth}
\author{Yu.~G.~Kolomensky}
\author{M.~J.~Lee}
\author{G.~Lynch}
\affiliation{Lawrence Berkeley National Laboratory and University of California, Berkeley, California 94720, USA }
% ...
\collaboration{The \babar\ Collaboration}
\noaffiliation


\begin{abstract}
We perform a measurement of the $\tau \rightarrow l \gamma \nu \bar \nu$ ($l= e, \mu$) 
branching fractions 
for a minimum photon energy of 10~MeV in the $\tau$ rest frame, 
using \lumi of \epem\ collisions collected at the center-of-mass energy of the $\Upsilon(4S)$ resonance 
with the \babar\ detector at the \pep2\ storage rings. 
We find $\mathcal B (\tau \rightarrow \mu \gamma \nu \bar \nu) = (3.69 \pm 0.03 \pm 0.10) \times 10^{-3}$, 
and $\mathcal B(\tau \rightarrow e \gamma \nu \bar \nu) = (1.847 \pm 0.015 \pm 0.052) \times 10^{-2}$, 
where the first quoted error is statistical, and the second is systematic. 
These results are substantially more precise than previous measurements. 
\end{abstract}


\maketitle


\begin{figure*}[!ht]
\centering
\includegraphics[width=0.8\textwidth]{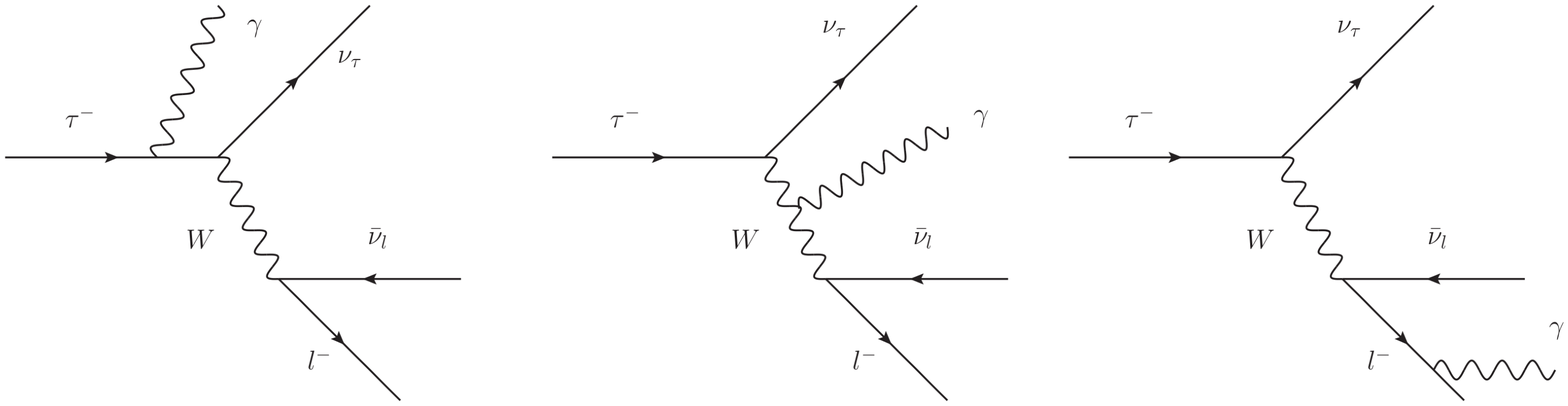}
\caption{Standard Model Feynman diagrams for $\tau \rightarrow l \gamma \nu \bar \nu$ at tree level.}
\label{sm_amp}
\end{figure*}

Leptonic $\tau$ decays are generally well suited to investigate the Lorentz structure of 
electroweak interactions in a model-independent way \cite{michel}. 
In particular, leptonic radiative decays $\tau \rightarrow l \gamma \nu \bar \nu$, where the 
charged lepton ($l$) is either an electron ($e$) or a muon ($\mu$), 
have been studied for a long time because they are sensitive 
to the anomalous magnetic moment of the $\tau$ lepton \cite{laursen}. 
At tree level, these decays can proceed through three Feynman diagrams depending on whether the 
photon is emitted by the incoming $\tau$, the outgoing charged lepton, 
or the intermediate $W$ boson, as shown in Fig.~\ref{sm_amp}. 
The amplitude for the emission of the photon by the intermediate 
boson is suppressed by a factor $(m_{\tau}/M_W)^2$ with respect to a photon from the incoming/outgoing 
fermions 
and is thus negligible with respect to next-to-leading order (NLO) QED radiative corrections \cite{theo}. 
Both branching fractions have been measured by the CLEO collaboration. 
CLEO obtained $\mathcal B (\tau \rightarrow \mu \gamma \nu \bar \nu) = (3.61 \pm 0.16 \pm 0.35) \times 10^{-3}$, 
and $\mathcal B (\tau \rightarrow e \gamma \nu \bar \nu) = (1.75 \pm 0.06 \pm 0.17) \times 10^{-2}$ 
for a minimum photon energy of 10~MeV in the $\tau$ rest frame \cite{cleo}. 
In addition, the OPAL collaboration finds 
$\mathcal B (\tau \rightarrow \mu \gamma \nu \bar \nu) = (3.0 \pm 0.4 \pm 0.5) \times 10^{-3}$ 
for a minimum photon energy of 20~MeV in the $\tau$ rest frame \cite{opal}. 
\newpage
In the present work we perform a measurement of $\tau \rightarrow l \gamma \nu \bar \nu$  
branching fractions 
for a minimum photon energy of 10~MeV in the $\tau$ rest frame. 
This analysis uses data recorded 
by the \babar\ detector at the \pep2\ asymmetric-energy \epem\ 
storage rings operated at the SLAC National Accelerator Laboratory.
The data sample consists of 431 fb$^{-1}$ of $e^+ e^-$ collisions recorded at 
at the center-of-mass energy (CM) $\sqrt{s} = 10.58 \gev$ \cite{lumi_paper}. 
The cross section for 
$\tau$-pair production is $\sigma_{\tau\tau} = 0.919\pm0.003$ nb
\cite{tautau} corresponding to a data sample of about $400\times 10^6$ $\tau$-pairs. 
A detailed description of the \babar\ detector is given elsewhere \cite{detector, det2}. 
Charged particle momenta are measured with a five-layer 
double-sided silicon vertex tracker and a 40-layer helium-isobutane 
drift chamber inside a 1.5 T superconducting solenoid magnet. 
An electromagnetic calorimeter (EMC) consisting of 6580 CsI(Tl) 
crystals is used to measure electron and photon energies; 
a ring-imaging Cherenkov detector is used to identify 
charged hadrons; 
the instrumented magnetic flux return (IFR) is used for muon identification. 
About half of the data were taken with the IFR embedded with resistive plate chambers, 
later partially replaced by limited streamer tubes. 

For this analysis, a Monte Carlo (MC) simulation 
is used to estimate the signal efficiency and to optimize the selection algorithm. 
Simulated $\tau$-pair events 
are generated using \kk2f \cite{kk}  
and $\tau$ decays are simulated with \tauola \cite{tauola}. 
Final-state radiative effects for $\tau$ decays in \tauola are simulated using \photos \cite{photos}. 
A signal $\tau$-pair MC sample is generated where one of the $\tau$ 
leptons decays to $\tau \rightarrow l \gamma \nu \bar \nu$, and the other 
decays according to known decay modes \cite{PDG}. 
For the signal sample we require the minimum photon energy in the $\tau$ rest frame 
to be $E^*_{\gamma, \mathrm{min}} > 10$~MeV. 
The $\tau \rightarrow l \gamma \nu \bar \nu$ decays with $E^*_{\gamma, \mathrm{min}} < 10$~MeV 
are treated as background. 
A separate $\tau$-pair MC sample is generated requiring each $\tau$ lepton to 
decay in a mode based on current experimental knowledge; we exclude signal events in 
the former sample to obtain a $\tau$-pair background sample. 
Other MC simulated background samples include $\mu^+ \mu^-$, $q \bar q$ ($u \bar u$, 
$d \bar d$, $s \bar s$, $c \bar c$), and $B \bar B$ ($B=B^+$, $B^0$) events. 
The $\mu^+ \mu^-$ events are generated by \kk2f , $q \bar q$ events are generated 
using the \jetset generator \cite{jetset} while $B \bar B$ events are simulated with \evtgen \cite{evtgen}.  
The detector response is simulated with \mbox{\tt GEANT4}~\cite{geant}. 
Background from two-photon and Bhabha events 
is estimated from data. 
\\
\indent 
The signature for $\tau \rightarrow l \gamma \nu \bar \nu$ decays is a charged particle (track), 
identified either as an electron or a muon, and an energy deposit (cluster) in the EMC 
not associated with any track, the photon. 
Since $\tau$ leptons decay mostly to a single charged particle, 
events with two well-reconstructed tracks and 
zero total charge are selected, where no track pair is 
consistent with being a photon conversion in the detector 
material. The transverse momentum of each track is required to be $p_T > 0.3$~GeV/c, 
the cosine of the polar angle is required to be between $-0.75$ and $0.95$ within the calorimeter acceptance range 
to ensure good particle identification. 
The total missing transverse moment of the event is required to be $p_{T,\mathrm{miss}} > 0.5$~GeV/$c$. 
All clusters in the EMC with no associated tracks (neutral clusters) 
are required to have a minimum energy of 50~MeV. 
We also reject events with neutral clusters having $E < 110$~MeV if they are within 25~cm of a track, where 
the distance is measured on the inner wall of the EMC. 

Each event is divided into hemispheres 
("signal" and "tag" hemispheres) in the CM frame by a plane 
perpendicular to the thrust axis, calculated using all reconstructed 
charged and neutral particles \cite{thrust}. 
For every event, the magnitude of the thrust is required to be between 0.9 and 0.995. 
The lower limit on the thrust magnitude rejects most $q \bar q$ events while the upper limit 
removes $e^+ e^- \rightarrow \mu^+ \mu^-$ and Bhabha events. 
The signal hemisphere must contain one track and one neutral cluster. 
The tag hemisphere must contain one track, identified 
either as an electron, muon or pion, and possibly one additional neutral cluster 
or $n \pi^0$ ($n=$1, 2). Each $\pi^0$ candidate is built up from a pair of neutral clusters 
with a di-photon invariant mass in the range $[100, 160]$~MeV. 
To further suppress di-muon and Bhabha events, we reject events 
where the leptons in the signal and tag hemispheres have the same flavor. 
Since there are at least three undetected neutrinos in the final state 
we require the total 
energy to be less than 9~GeV. 
In the signal hemisphere, we require that the 
distance ($d_{l \gamma}$) between the track and the neutral cluster, 
measured on the inner wall of the EMC, to be less than 100~cm. 

Electrons are identified by applying an 
Error Correcting Output Code (ECOC) \cite{ecoc} algorithm based on 
Bagged Decision Tree (BDT) \cite{BDT} classifiers 
using as input the 
ratio of the energy in the EMC 
to the magnitude of the momentum of the track $(E/p)$, the ionization 
loss in the tracking system $(\dedx)$, and the shape of the shower in the electromagnetic calorimeter. 

Muon identification makes use of a BDT 
algorithm, 
using as input the number of hits in the IFR, the number of interaction lengths traversed,
and the energy deposition in the calorimeter. Since muons with momenta less than $500\mevc$ do not 
penetrate into the IFR, the BDT also uses information 
the energy loss $dE/dx$ in the tracking system 
to maintain a very low $\pi-\mu$ misidentification probability with high selection efficiencies.
The electron and muon identification efficiencies are 91\% and 62\%, respectively. 
The probability for a $\pi$ to be misidentified as an $e$ 
is below 0.1\%, while the probability to be misidentified 
as a $\mu$ is around 1\% depending on momentum. 

\begin{figure*}[!ht]
\centering
\includegraphics[width=0.4456\textwidth]{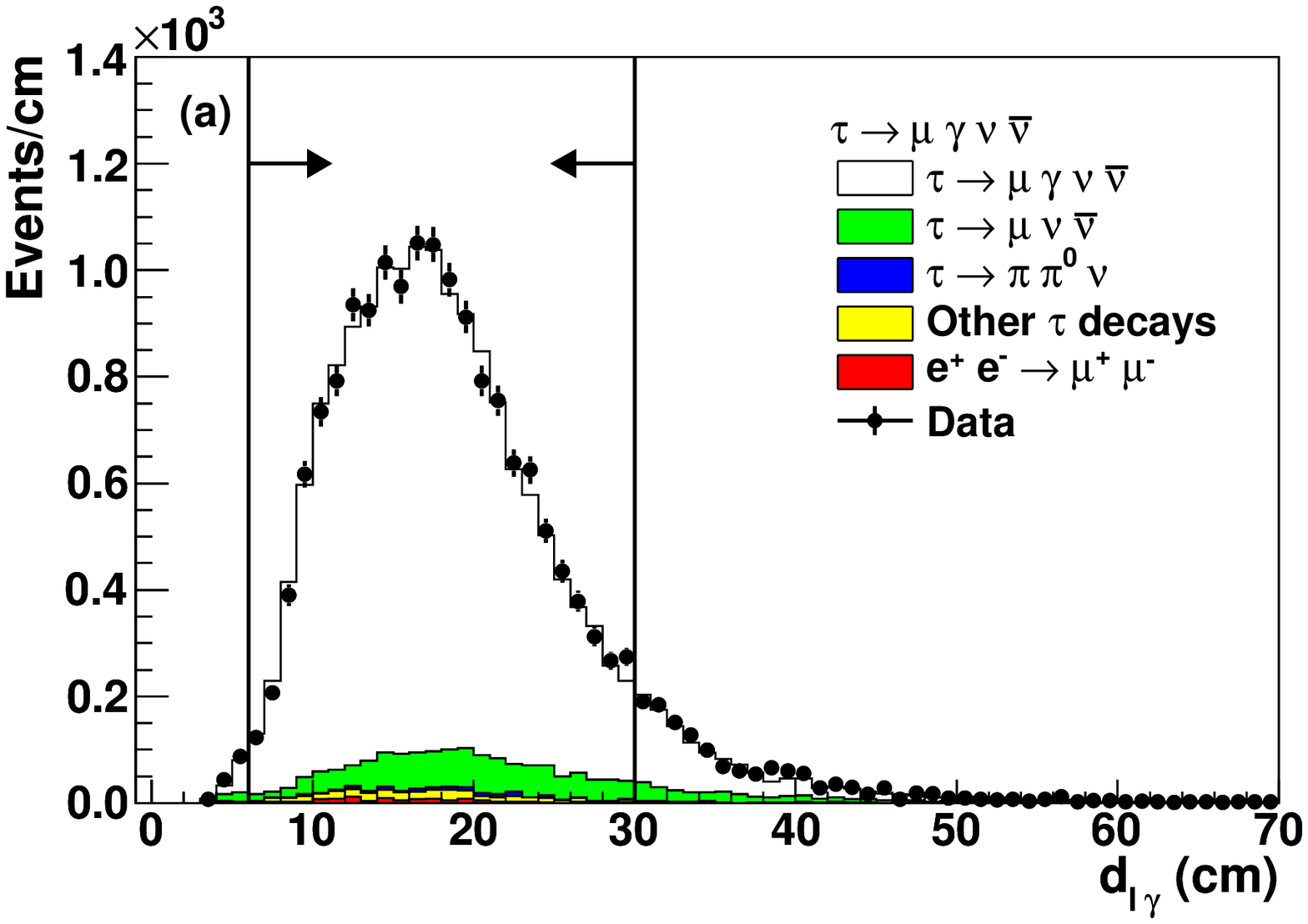}
\quad
\includegraphics[width=0.4456\textwidth]{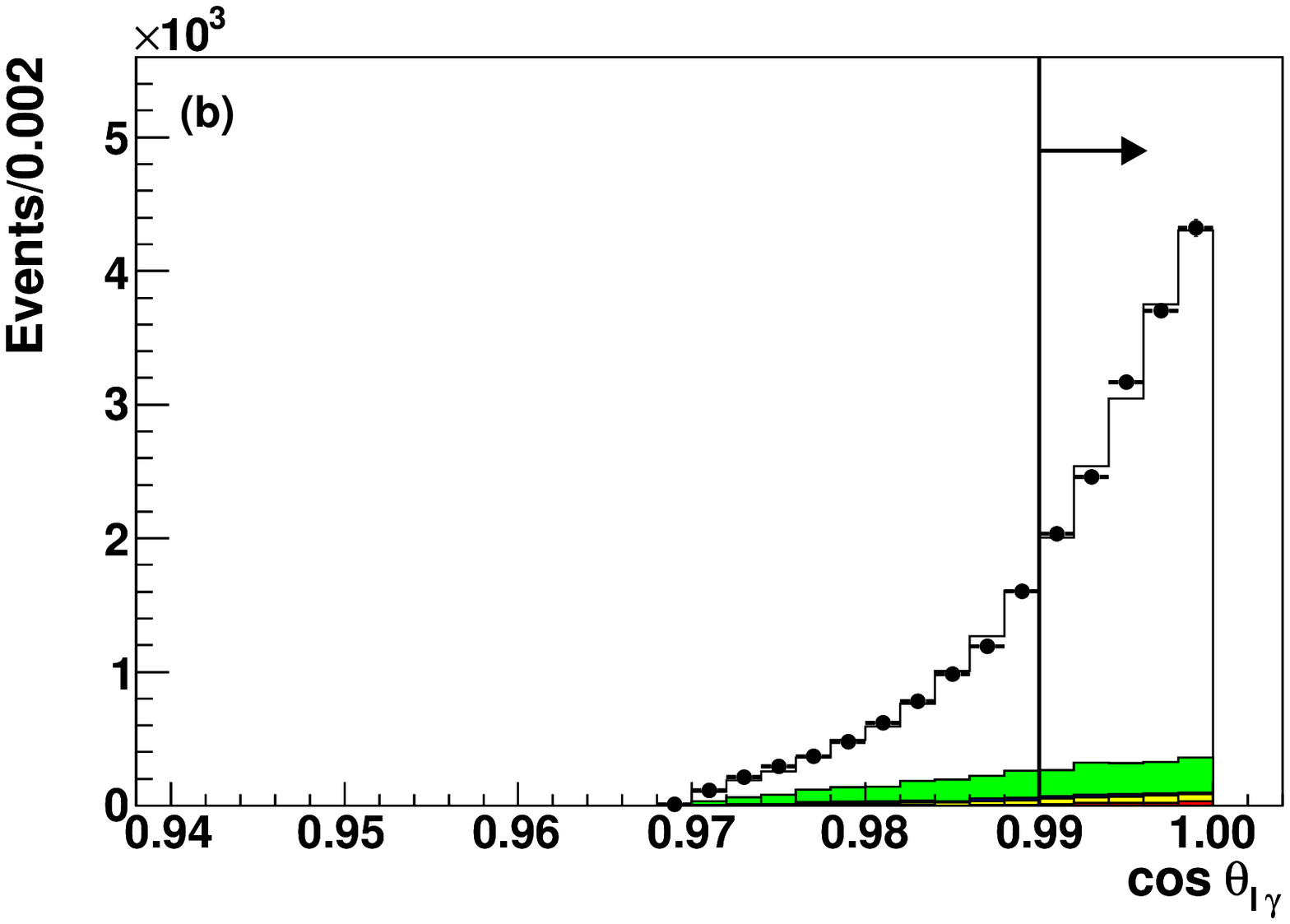}
\quad
\includegraphics[width=0.4456\textwidth]{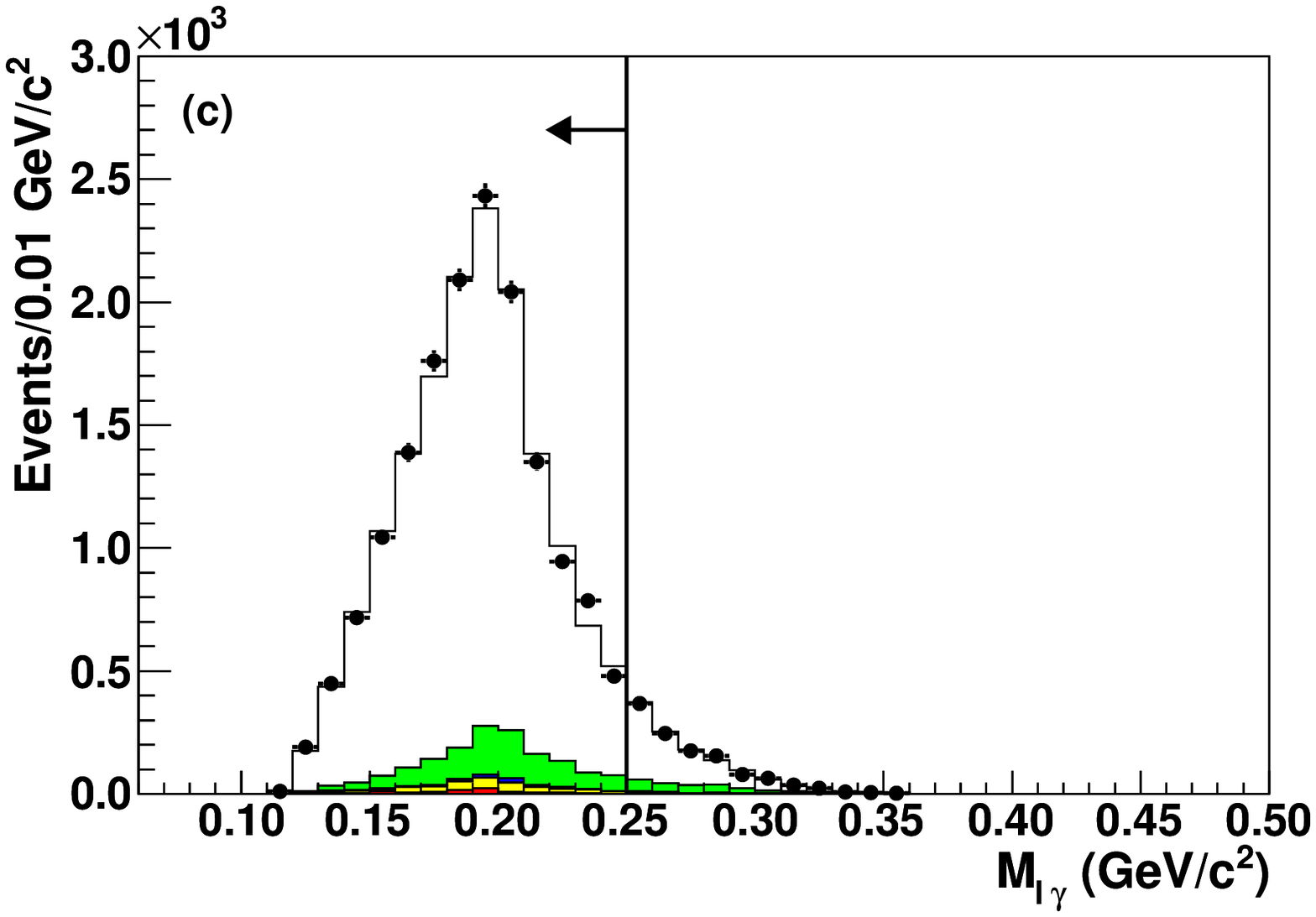}
\quad
\includegraphics[width=0.4456\textwidth]{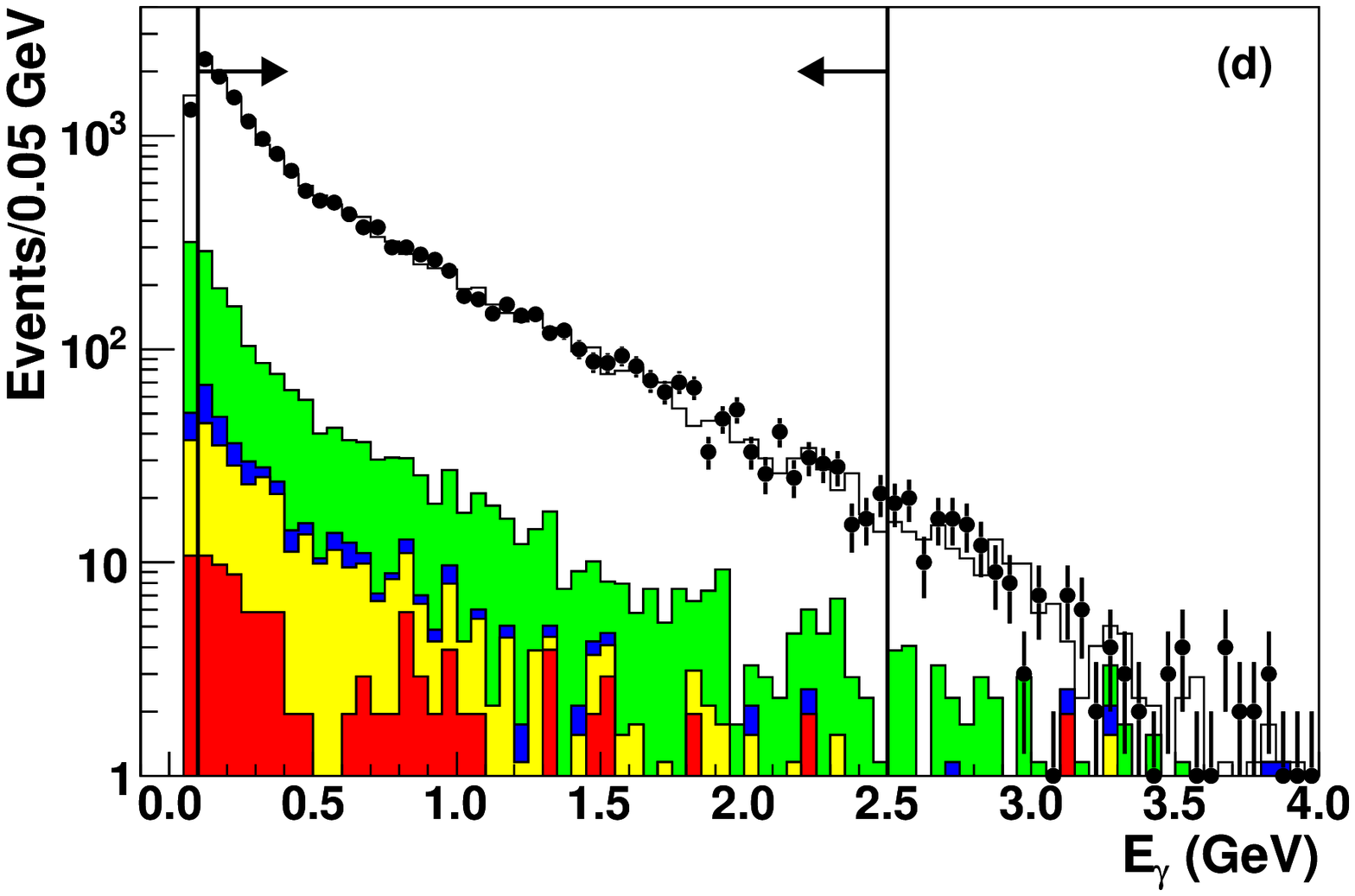}
\caption{Selection of the $\tau \rightarrow \mu \gamma \nu \bar \nu$: 
(a) distance between lepton and photon candidates on the inner EMC wall, 
(b) cosine of the angle between momenta of the lepton and photon candidates in the CM frame, 
(c) invariant mass of the lepton photon pair, and (d) photon candidate energy in the CM frame 
for radiative $\tau$ decay into a muon after applying all selection criteria 
except the one on the plotted quantities. The selection criteria on the plotted quantities 
are highlighted by the vertical lines; we retain the regions indicated by the horizontal arrows. 
}
\label{fig1}
\end{figure*}

\begin{figure*}[!htb]
\centering
\includegraphics[width=0.4456\textwidth]{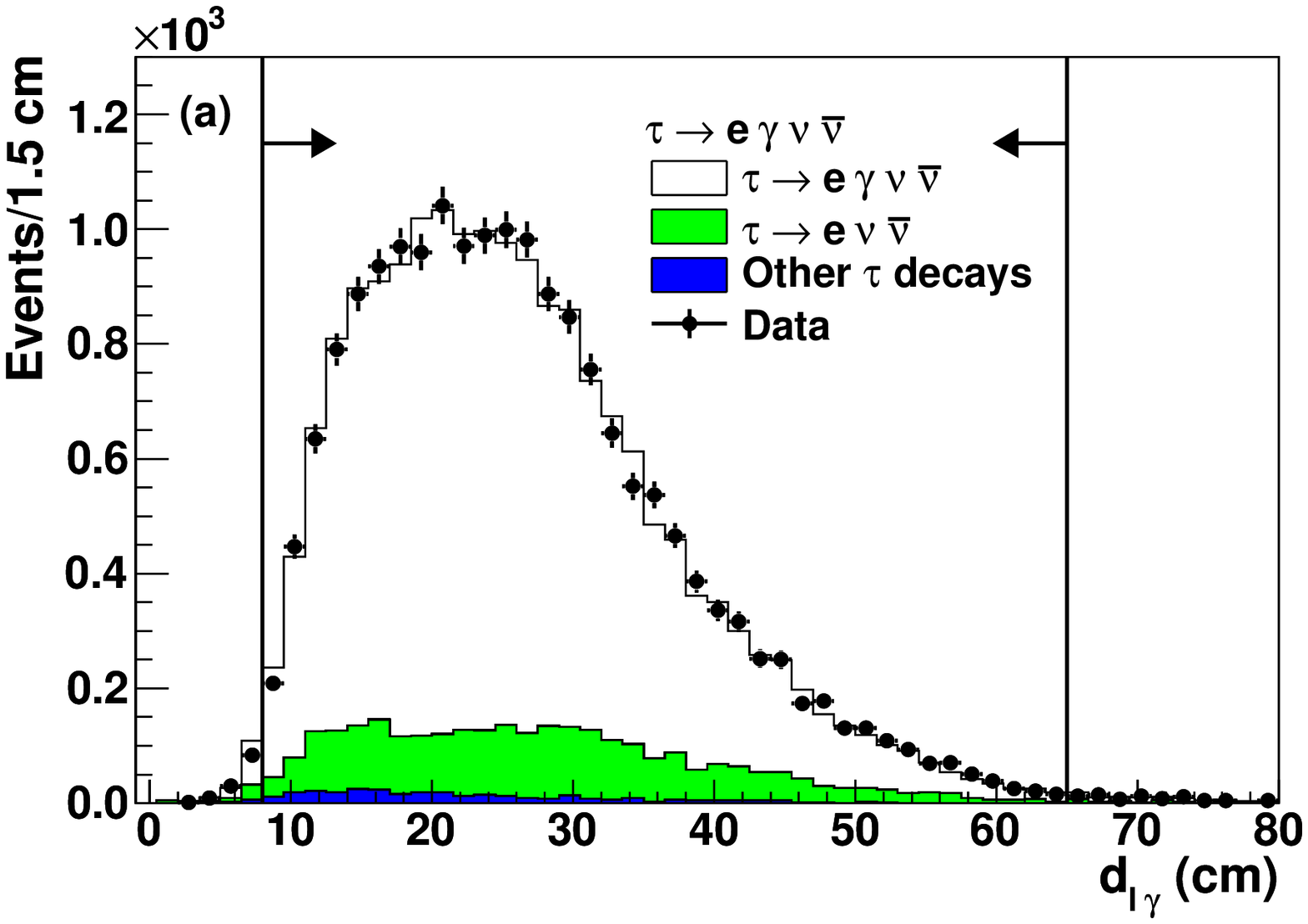}
\quad
\includegraphics[width=0.4456\textwidth]{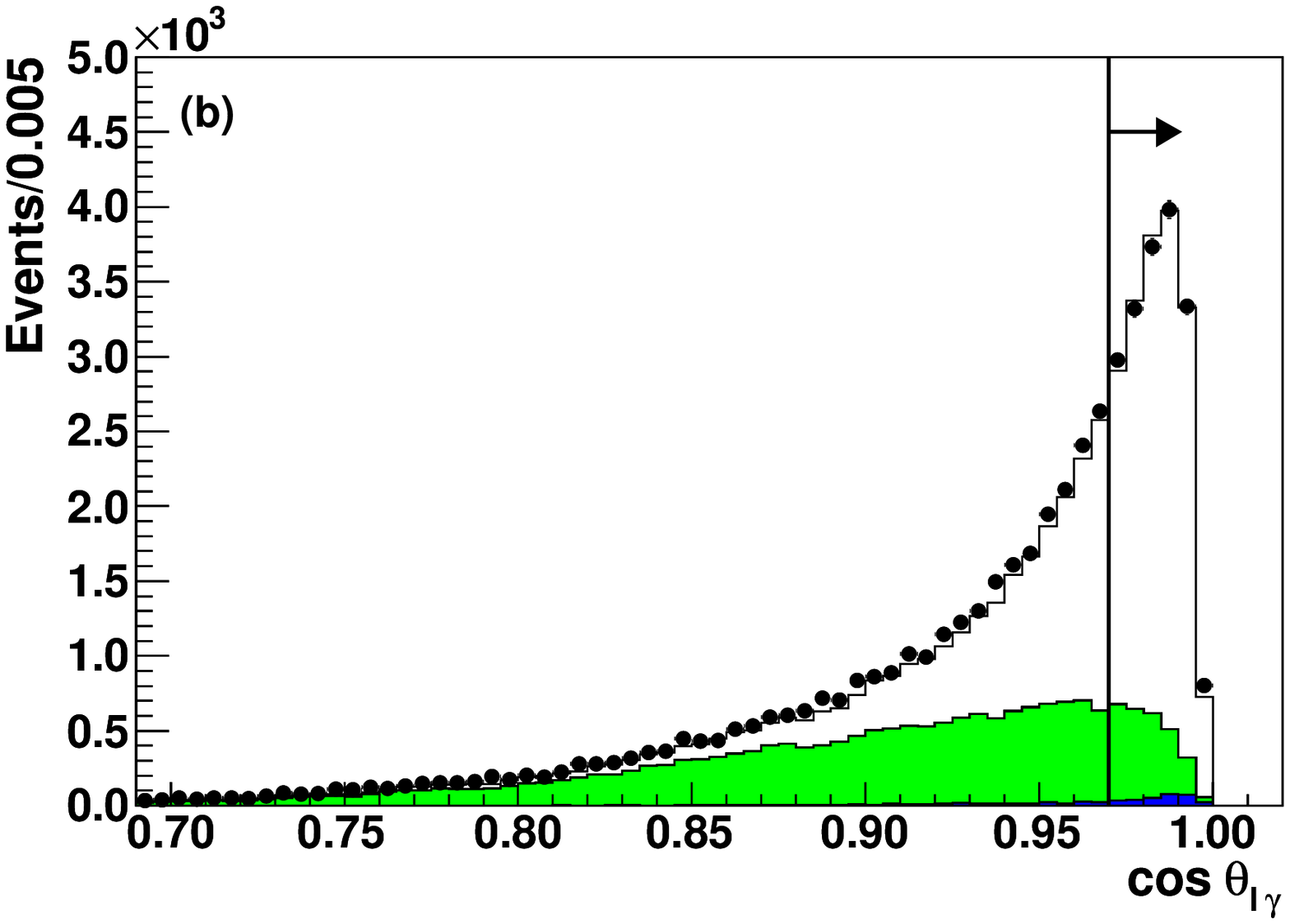}
\quad
\includegraphics[width=0.4456\textwidth]{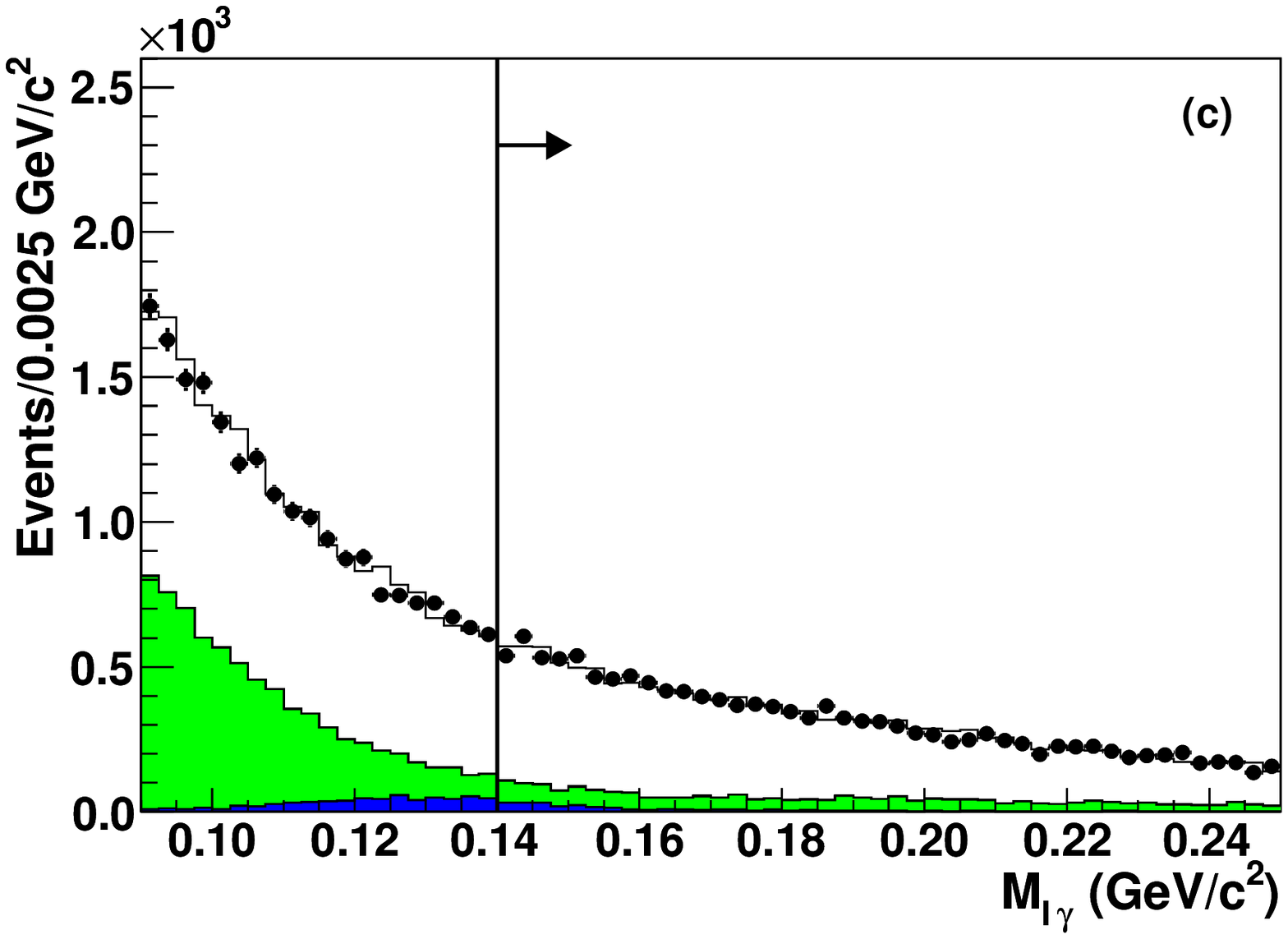}
\quad
\includegraphics[width=0.4456\textwidth]{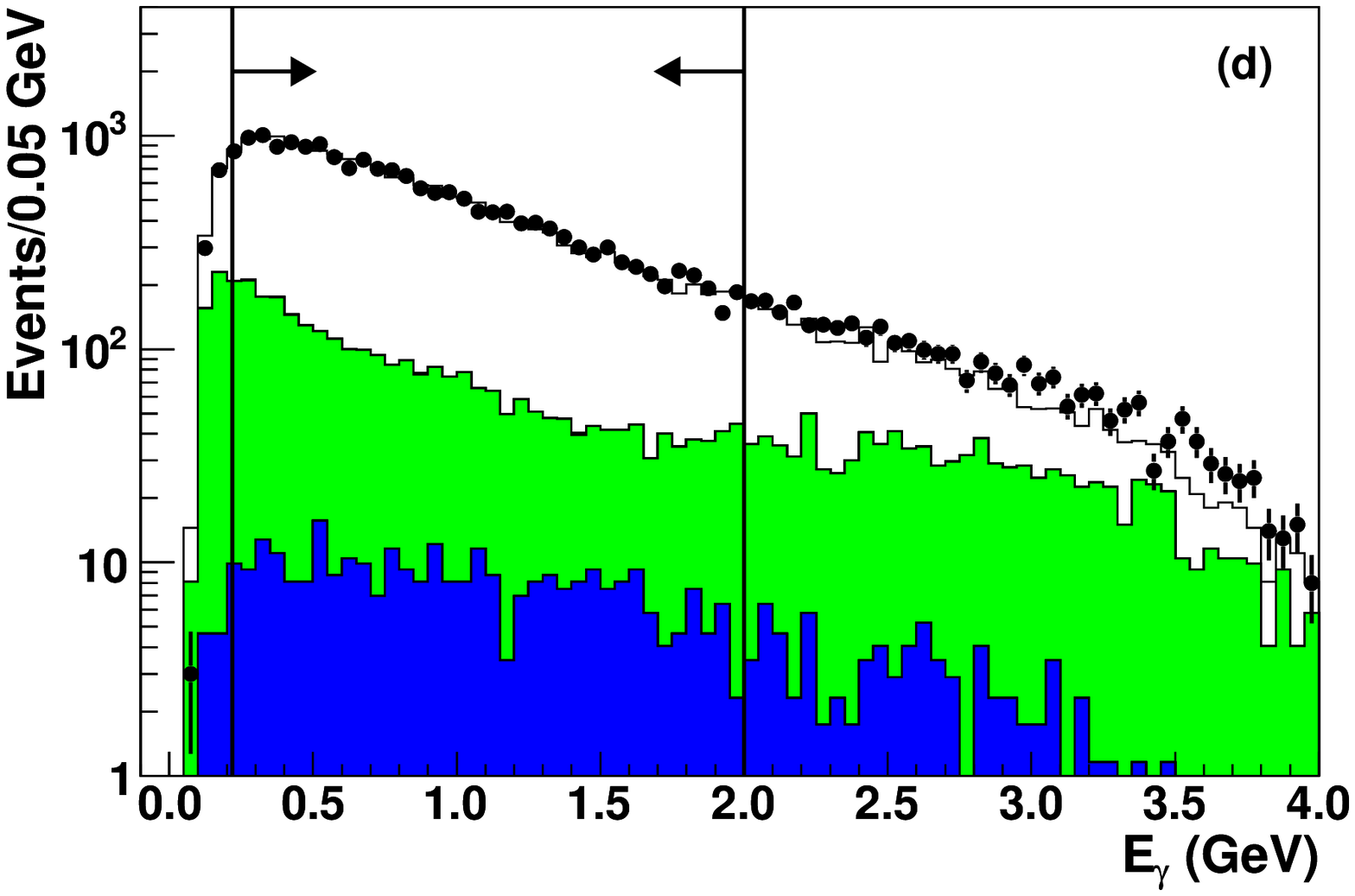}
\caption{Selection of the $\tau \rightarrow e \gamma \nu \bar \nu$ sample: 
(a) distance between lepton and photon candidates on the inner EMC wall, 
(b) cosine of the angle between momenta of the lepton and photon candidates in the CM frame, 
(c) invariant mass of the lepton photon pair, and (d) photon candidate energy in the CM frame 
for radiative $\tau$ decay into an electron after applying all selection criteria 
except the one on the plotted quantities. The selection criteria on the plotted quantities 
are highlighted by the vertical lines; we retain the regions indicated by the horizontal arrows. 
}
\label{fig2}
\end{figure*}

After the preselection, both samples are dominated by background events. 
For the $\tau \rightarrow \mu \gamma \nu \bar \nu$ sample, the main background sources 
are initial-state radiation (ISR), $\tau \rightarrow \pi \pi^0 \nu$ decays, 
$e^+ e^- \rightarrow \mu^+ \mu^-$ events, 
and $\tau \rightarrow \pi \nu$ decays. 
For the $\tau \rightarrow e \gamma \nu \bar \nu$ sample, almost all background contribution is from 
$\tau \rightarrow e \nu \bar \nu$ decays in which the electron radiates a photon 
in the magnetic field of the detector (bremsstrahlung). 
Further background suppression is obtained by placing requirements on the angle between 
the lepton and photon in the CM frame ($\cos \theta_{l \gamma}$). 
For $\tau \rightarrow \mu \gamma \nu \bar \nu$ we require $\cos \theta_{l \gamma} > 0.99$, 
while for $\tau \rightarrow e \gamma \nu \bar \nu$ we require $\cos \theta_{l \gamma} > 0.97$ 
(see Figs. \ref{fig1} and \ref{fig2}). 
To reject background from $\tau \rightarrow e \nu \bar \nu$ decays 
in the $\tau \rightarrow e \gamma \nu \bar \nu$ sample, we further impose 
a minimum value for the invariant mass of the lepton-photon pair $M_{l \gamma} \ge 0.14$~GeV/$c^2$ for this channel. 
In addition to the aforementioned quantities, 
the selection criteria use the energy of the photon and $d_{l \gamma}$. 
The selection criteria are optimized in order to give the smallest 
statistical and systematic uncertainty on the branching fractions. 

After optimization, for $\tau \rightarrow \mu \gamma \nu \bar \nu$, we require 
$\cos \theta_{l \gamma} \ge 0.99$, $ 0.10 \le E_{\gamma} \le 2.5$~GeV, $ 6 \le d_{l \gamma} \le 30$~cm, 
and $M_{l \gamma} \le 0.25$~GeV/$c^2$. The requirement on $M_{l \gamma}$ rejects 
backgrounds from non-signal $\tau$ decays. 
For the $\tau \rightarrow e \gamma \nu \bar \nu$ channel, we require 
$\cos \theta_{l \gamma} \ge 0.97$, $ 0.22 \le E_{\gamma} \le 2.0$~GeV, $8 \le d_{l \gamma} \le 65$~cm 
in addition to $M_{l \gamma} \ge 0.14$~GeV/$c^2$. 

The signal efficiencies, the fraction of background events, and the 
number of events selected in the data are given in Table~\ref{tab:results}. 

\begin{table}[!htb]
\begin{ruledtabular}
\caption{Signal efficiencies $\epsilon$ (\%), expected fractional background contribution $f_{\mathrm{bkg}} = 
N_{\mathrm{bkg}}/(N_{\mathrm{sig}} + N_{\mathrm{bkg}})$, 
where $N_{\mathrm{sig}}$ is the number of signal events and $N_{\mathrm{bkg}}$ is the number of background events, and 
number of observed events ($N_{\mathrm{obs}}$) 
for the two decay modes after applying all selection criteria. All quoted uncertainties are statistical.}
\label{tab:results}
\begin{tabular}{lcc}
& $\tau \rightarrow \mu \gamma \nu \bar \nu$ & $\tau \rightarrow e \gamma \nu \bar \nu$ \\
\hline
$\epsilon$ &  $0.480 \pm 0.010 $  & $0.105 \pm 0.003 $  \\
$f_{\mathrm{bkg}}$ &  $0.102 \pm 0.002$  &  $0.156 \pm 0.003$ \\
$N_{\mathrm{obs}}$ &  $ 15688 \pm 125 $ &  $ 18149 \pm 135 $ \\
\end{tabular}
\end{ruledtabular}
\end{table}

The branching fraction is determined using
\begin{equation}
\mathcal B_l = \frac{N_{\mathrm{obs}} 
(1 - f_{\mathrm{bkg}})}{2 \mbox{ } \sigma_{\tau \tau} \mbox{ } \mathcal L \mbox{ } \epsilon}
\nonumber
\end{equation}
where 
\begin{linenomath} 
$N_{\mathrm{obs}}$ 
\end{linenomath}
is the number of observed events, 
$\sigma_{\tau \tau}$ is the cross section for $\tau$ pair production, $\mathcal L$ is the total integrated luminosity, 
and the signal efficiency $\epsilon$ is determined from the MC sample. 

After applying all selection criteria, we find 
\begin{eqnarray*}
\mathcal B(\tau \rightarrow \mu \gamma \nu \bar \nu) & = & (3.69 \pm 0.03 \pm 0.10) \times 10^{-3} \\
\mathcal B(\tau \rightarrow e \gamma \nu \bar \nu) & = & (1.847 \pm 0.015 \pm 0.052) \times 10^{-2}
\end{eqnarray*} 
where the first error is statistical and the second is systematic. 
The systematic uncertainties on signal efficiency and on the number of the expected background 
events affect the final result, and are 
summarized in Table \ref{tab:syst}. 
The most important contributions to the total uncertainty are from 
the uncertainties on particle identification, and photon detection efficiency. 

To estimate the uncertainty on photon detection efficiency, we rely on 
$e^+ e^- \rightarrow \mu^+ \mu^- \gamma$ events for the high energy region ($E_{\gamma} > 1$~GeV), 
and photons from $\pi^0$ decays for the low energy region ($E_{\gamma} < 1$~GeV). 
Using fully reconstructed $e^+ e^- \rightarrow \mu^+ \mu^- \gamma$ 
events, we find that the photon detection efficiency for data and MC samples are consistent within 1\% for $E_{\gamma}>1$~GeV. 
For photon energies $E_{\gamma}<1$~GeV, we measure 
the ratio of the branching fractions 
for $\tau \rightarrow \pi \nu$ and $\tau \rightarrow \rho \nu$ decays. 
The resulting uncertainty on the $\pi^0$ reconstruction efficiency is found to be below 3\%. 
Taking into account the 1.1\% uncertainty on the branching fractions, 
the resulting energy-averaged uncertainty on the single photon detection efficiency is 1.8\%. 
We use this value as the systematic uncertainty in the efficiency 
for $\tau \rightarrow l \gamma \nu \bar \nu$. 

The uncertainties on particle identification efficiency 
are estimated using control samples, by measuring the deviation of the data and MC efficiencies for 
tracks with the same kinematic properties. 
The uncertainty on the efficiency of the electron identification is 
evaluated using a control sample consisting of radiative and non-radiative Bhabha events, while the uncertainty 
for muons is an $e^+ e^- \rightarrow \mu^+ \mu^- \gamma$ control sample. 
The uncertainty on the probability of misidentifying the pion as a muon or electron is evaluated 
using samples of $\tau \rightarrow \pi \pi \pi \nu$ decays. 
The corresponding systematic uncertainty on the efficiency 
for $\tau \rightarrow l \gamma \nu \bar \nu$ is 1.5\% for both channels. 

For the background estimation, we define control regions that are enhanced with background events. 
For $\tau \rightarrow \mu \gamma \nu \bar \nu$, where the major background contribution 
is not peaking in $\cos \theta_{\mu \gamma}$, 
we invert the cut on $\cos \theta_{\mu \gamma}$. 
For $\cos \theta_{\mu \gamma} < 0.8$,  the maximum expected signal rate 
is 3\% of the corresponding background rate. 
The maximum discrepancy between the MC sample prediction and the number of observed events is 8\%, 
with an excess of events in the MC sample. 
We take this discrepancy as estimate of the uncertainty on the background prediction.
For $\tau \rightarrow e \gamma \nu \bar \nu$, where the major background contributions have similar 
$\cos \theta_{e \gamma}$ distributions as signal, we apply a similar strategy after 
requiring the invariant mass $M_{l \gamma} < 0.14$~GeV/$c^2$; in this case 
we take $\cos \theta_{e \gamma} < 0.90$. The maximum contamination of signal events in this region 
is 10\%, and the maximum discrepancy between the prediction and the number of observed events is 4\% 
with an excess of data events. We take this value as an estimate of the uncertainty on the background rate. 
The errors on the branching fractions due to the uncertainty on background estimates 
are 0.9\% for $\tau \rightarrow \mu \gamma \nu \bar \nu$, and 0.7\% for $\tau \rightarrow e \gamma \nu \bar \nu$, 
respectively (Table~\ref{tab:syst}). 
Cross-checks of the background estimation are performed by considering 
the number of events expected and observed in different sideband regions immediately 
neighboring the signal region for each decay mode and found to be compatible 
with the aforementioned systematic uncertainties. 

All other sources of uncertainty, including current 
knowledge of the $\tau$ branching fractions \cite{PDG} (BF), total number of $\tau$ pairs, 
limited MC statistics, dependence on selection criteria, 
and track momentum resolution are found to be 
smaller than $1.0 \%$. 

In conclusion, we have made a measurement of the branching fractions of the radiative leptonic $\tau$ 
decays $\tau \rightarrow e \gamma \nu \bar \nu$ and $\tau \rightarrow \mu \gamma \nu \bar \nu$, 
for a minimum photon energy of 10~MeV in the $\tau$ rest frame, 
using the full dataset of \epem\ collisions collected 
by \babar\ at the center-of-mass energy of the $\Upsilon(4S)$ resonance. 
We find $\mathcal B (\tau \rightarrow \mu \gamma \nu \bar \nu) = (3.69 \pm 0.03 \pm 0.10) \times 10^{-3}$, 
and $\mathcal B(\tau \rightarrow e \gamma \nu \bar \nu) = (1.847 \pm 0.015 \pm 0.052) \times 10^{-2}$, 
where the first error is statistical and the second is systematic. 
These results are more precise by a factor of three compared to previous experimental measurements. 
Our results are in agreement with the Standard Model values at tree level, 
$\mathcal B (\tau \rightarrow \mu \gamma \nu \bar \nu) = 3.67 \times 10^{-3}$, 
and $\mathcal B(\tau \rightarrow e \gamma \nu \bar \nu) = 1.84 \times 10^{-2}$ \cite{theo}, 
and with current experimental bounds. 
\vadjust{\penalty-10000} 
We are grateful for the extraordinary contributions of our \pep2\ colleagues in 
achieving the excellent luminosity and machine conditions 
that have made this work possible. 
The success of this project also relies critically on the 
expertise and dedication of the computing organizations that 
support \babar. 
The collaborating institutions wish to thank 
SLAC for its support and the kind hospitality extended to them. 
This work is supported by the US Department of Energy 
and National Science Foundation, the Natural Sciences and Engineering Research Council (Canada), 
the Commissariat \`a l'Energie Atomique and Institut National de Physique Nucl\'eaire et de Physique des Particules 
(France), the Bundesministerium f\"ur Bildung und Forschung and 
Deutsche Forschungsgemeinschaft (Germany), the Istituto Nazionale di Fisica Nucleare (Italy), 
the Foundation for Fundamental Research on Matter (The Netherlands), 
the Research Council of Norway, the 
Ministry of Education and Science of the Russian Federation, 
Ministerio de Econom\'{\i}a y Competitividad (Spain), the 
Science and Technology Facilities Council (United Kingdom), 
and the Binational Science Foundation (U.S.-Israel). 
Individuals have received support from 
the Marie-Curie IEF program (European Union) and the A. P. Sloan Foundation (USA).

\begin{table}[htb]
\begin{ruledtabular}
\caption{Summary of systematic contributions (\%) to the branching fraction 
from the different uncertainty sources for the two signal channels. 
The total systematic uncertainties are obtained summing in quadrature 
the various systematic uncertainties for each decay channel.}
\label{tab:syst}
\begin{tabular}{lcc}
& $\tau \rightarrow \mu \gamma \nu \bar \nu$ & $\tau \rightarrow e \gamma \nu \bar \nu$\\
\hline
Photon efficiency & 1.8  & 1.8 \\
Particle identification & 1.5  & 1.5 \\
Background evaluation & 0.9  & 0.7 \\
BF \cite{PDG} & 0.7  & 0.7 \\
Luminosity and cross section & 0.6  & 0.6 \\
MC statistics & 0.5  & 0.6 \\
Selection criteria & 0.5  & 0.5  \\
Trigger selection & 0.5  & 0.6 \\
Track reconstruction & 0.3  & 0.3 \\
\hline
Total & 2.8  & 2.8 \\
\end{tabular}
\end{ruledtabular}
\end{table}

\end{document}